\documentstyle[aps,prb,twocolumn]{revtex}
\begin{document}
\title{Ergodic versus nonergodic behavior in oxygen
deficient high-$T_c$ superconductors}
\author{Sergei A. Sergeenkov}
\address{Bogoliubov Laboratory of Theoretical Physics,
Joint Institute for Nuclear Research,\\ 141980 Dubna, Moscow region, Russia}
\address{
\centering{
\medskip\em
\begin{minipage}{14cm}
{}~~~
The oxygen defects induced phase transition from nonergodic to
ergodic state in superconductors with intragrain granularity is considered
within the superconductive glass model. The model predictions are found to
be in a qualitative agreement with some experimental observations in
deoxygenated high-T$_c$ single crystals.
{}~\\
\medskip
{}~
\end{minipage}
}}
\maketitle
\narrowtext

\section{Introduction}

According to the recent findings (see, e.g., [1-8] and references
therein), high-$T_c$ superconductors (HTS) exhibit an
anomalous (nonclassical [8]) magnetic-field behavior, which has been attributed to the
''field-induced intragrain granularity'' in oxygen deficient samples and
interpreted in terms of the field-induced decoupling of regions of
oxygen-rich material by boundaries of oxygen-poor material. A ''phase
diagram'' $H_m(\delta ,T)$, that demarcates the multigrain onset as a
function of temperature and oxygen deficiency,$\delta $, was found [3] to
confirm that oxygen-deficient single crystals exhibit behavior
characteristic of homogeneous superconductors for $H<H_m$ and inhomogeneous
superconductors for $H>H_m$. The granular behavior for $H>H_m$ has been
related to the clusters of oxygen defects (within the CuO plane) that
restrict supercurrent flow and allow excess flux to enter the crystal. The
observed $H_m(\delta ,T)$ data were described by a 2-D percolation model for
oxygen defects. It means that there exists a critical oxygen deficiency,$%
\delta _c$, above which there are no continuous current paths. For $\delta $
greater than $\delta _c$, oxygen-rich superconducting ''grains'' are
separated by oxygen-poor insulating boundaries so that there is no
superconducting path through a sample. For $\delta $ less than $\delta _c$,
a complete current path spans the sample and resistance measurements show
metallic behavior with a superconducting transition. Since $H_m(\delta ,T)$
signals the onset of granularity, a sample with $H_m(\delta ,T)=0$ implies
that the crystal has so many oxygen defects that it never exhibits
single-grain behavior [3].

The aim of the present paper is to show how the lack of oxygen in
HTS materials inspires the phase transition from nonergodic (in nearly
full-oxygenated crystals) to ergodic (in highly
oxygen-depleted crystals) state within the
so-called superconductive glass (SG) model (see, e.g., [9-15] and references
therein), and to face the model predictions with some experimental data for
deoxygenated HTS single crystals. More specifically, the nonergodic (phase-
coherent) state is attributed to nontrivial equilibrium (long-time) behavior
of the defect-free crystal (with $\delta \simeq 0$) while
the ergodic (paracoherent) phase corresponds to the equilibrium state of
the defected crystal (when $\delta \simeq \delta _c$).

\section{The model}

The SG model is based on the well-known Hamiltonian of a granular
superconductor which in the so-called pseudospin representation has the
form [9-15]
\begin{eqnarray}
{\cal H}_0 &=&-\sum_{ij}^NJ(\delta ,T)\cos \phi _{ij}(\vec H)\\ \nonumber
&\equiv &-Re\left\{\sum_{ij}^NJ_{ij}S_i^{+}S_j^{-}\right\},
\end{eqnarray}
where
\begin{eqnarray}
J_{ij}(\delta ,T,\vec H)&=&J(\delta ,T)\exp [iA_{ij}(\vec H)],\\ \nonumber
\phi _{ij}(\vec H)&=&\phi _i-\phi _j-A_{ij}(\vec H),
\end{eqnarray}
\begin{eqnarray}
A_{ij}(\vec H)&=&\frac \pi {\phi _0}(\vec H\times \vec R_{ij})\vec r_{ij},\\ \nonumber
\vec r_{ij}&=&\vec r_i-\vec r_j,\qquad \vec R_{ij}=(\vec r_i+\vec r_j)/2.
\end{eqnarray}
This model describes the infinite-range interaction between oxygen-rich
superconducting grains [with phase $\phi _i(t)$ or Josephson pseudospins $%
S_i^{+}=e^{+i\phi _i}$], arranged in a random two-dimensional (2D)
lattice (modeling the CuO plane of oxygen-depleted $YBa_2Cu_3O_{7-\delta }$,
where a glass-like picture is established [1-7]) with coordinates $\vec
r_i=(x_i,y_i,0)$. The grains are separated by oxygen-poor insulating
boundaries producing Josephson coupling with energy $J(\delta ,T)$. The
system is under the influence of a frustrating applied magnetic field $\vec
H $,which is assumed to be normal to the CuO plane of HTS. The increase of
the oxygen deficiency, $\delta $,leads to the decrease of the Josephson
energy (via the increase of the insulating layer between oxygen-rich
grains). For small $\delta $ (such that $\delta \ll 1$) we can approximate
the $\delta $ dependence of the Josephson energy by a linear law [4], namely
$J(\delta ,T)\approx J(0,T)(1-\delta )$. The superconducting current through
the Josephson junction between grains i and j
\begin{eqnarray}
I_{ij}^s(\vec H)&=&\frac{2eJ}\hbar \sin \phi _{ij}(\vec H)\\ \nonumber
&\equiv &\frac{2e}\hbar Im\left\{ J_{ij}S_i^{+}S_j^{-}\right\}
\end{eqnarray}
induces a diamagnetic moment of the weak-link network [9-13]
\begin{equation}
\vec \mu =\pi \sum_{ij}^NI_{ij}^s(\vec H)(\vec r_{ij}\times \vec R_{ij})
\end{equation}
To study dynamic (relaxation) behavior of the model (1), let us
assume that in addition to the constant frustrating field $\vec H$, the
superconducting grains are under the influence of a small time-varying field
$\vec H_1(t)\ll \vec H$, so that $\cos (\phi _i-\phi _j-A_{ij}(\vec H+\vec
H_1(t)))\cong \cos \phi _{ij}(\vec H)+A_{ij}(\vec H_1(t))\sin \phi
_{ij}(\vec H)$. In view of Eqs.(1)-(5), the total (perturbed) Hamiltonian
can be cast into the form
\begin{equation}
{\cal H}(t)={\cal H}_0(\vec H)-\vec \mu \vec H_1(t)
\end{equation}
If the perturbation is applied continuously from $t=-\infty $ up to $t=0$
and is cut off at $t=0$, then the linear (with respect to the small
perturbation field $\vec H_1(t)=\vec H_1\theta (-t)$) response $M(t)\equiv
\overline{<{\mu _z}>}/V$ will relax to its equilibrium value $M_{eq}\equiv
\lim _{t\to \infty }M(t)$ according to the formula [17]
\begin{eqnarray}
M(t)-M_{eq}&=&\int\limits_{-\infty }^tdt^{\prime }G(t-t^{\prime })\vec
H_1(t^{\prime })\\ \nonumber
&=&\int\limits_t^\infty dt^{\prime }G(t^{\prime })\vec H_1
\end{eqnarray}
According to the fluctuation-dissipation theorem [17], the response function $%
G(t)$ is related to the relaxation function $\Phi (t)$ as follows, $%
G(t)=-(\partial /\partial t)\Phi (t)$. Thus the above equation reads
\begin{equation}
M(t)-M_{eq}=\frac 1VH_1[\Phi (t)-\Phi (\infty )],
\end{equation}
where
\begin{equation}
\Phi (t)=\beta \overline{<{\mu _z(t)}{\mu _z(0)}>}.
\end{equation}
Here $\beta =1/k_BT$, the bar denotes the configurational averaging over the
randomly distributed grain coordinates (see Appendix A), $<\ldots >$ means
the thermodynamic averaging with the Hamiltonian ${\cal H}_0(\vec H)$, and
we have assumed that $\vec H=(0,0,H)$ and $\vec H_1=(0,0,H_1)$. Therefore,
the function $\Phi (t)$ describes the relaxation of magnetization $M(t)$
after removal of the outer disturbance. As a result of configurational
averaging, the relaxation of magnetization can be approximated by the
formula (see Appendix A)
\begin{eqnarray}
M(t)&=&M(\delta ,T,H,H_1)\left| D(t)\right| ^2,\\ \nonumber
M(\delta ,T,H,H_1)&=&\chi (\delta ,T,H)H_1,
\end{eqnarray}
where
\begin{equation}
\chi (\delta ,T,H)\equiv \frac{16e^2s^2N^2J^2(\delta ,T)}{k_BTV\hbar ^2}
\left( \frac H{H_0}\right)^2\left(1+\frac{H^2}{H_0^2}\right)^{-4}
\end{equation}
Here $H_0={\phi _0}/s$ is a characteristic Josephson field with $s=\pi d^2$
an average JJ projection area, and N is the number of grains. Thus,all
information about the dynamic (relaxation) properties of the system is
contained in the time-dependent correlator $D(t)\equiv
(1/N)\sum_{ij}D_{ij}(t)$ which is defined as follows (see Appendix A)
\begin{equation}
D_{ij}(t)=\overline{<S_i^{+}(t)S_j^{-}(0)>}
\end{equation}

\section{Discussion}

As is well-known [18], there can be many types of long-time behavior
of $D(t)$. Two of them are of particular interest:
\begin{equation}
\lim _{t\to \infty }D(t)={L(T,H)\neq 0\qquad (I)}
\end{equation}
and
\begin{equation}
\lim _{t\to \infty }D(t)\propto {exp(-t/\tau )\ \qquad \ (II)}
\end{equation}
A simple example of the system belonging to class I (so-called nonergodic
state) has been discussed by de Gennes and Tinkham [19]. They considered the
long-time behavior of $D(t)$ for a superconducting thin film (of thickness $a
$) with diffuse reflection on the boundaries, in a parallel magnetic field.
When no volume defects are taken into account (pure limit), the system was
found to exhibit a nonergodic behavior with field-dependent nonergodicity
parameter $L(T,H)$, namely
\begin{equation}
L(T,H)\cong {\left\{ {1-(\pi Ha^2/3\phi _0),\qquad H\ll (\phi _0/a^2)} \atop
{(2\phi _0/\pi Ha^2)^2,\qquad H\gg (\phi _0/a^2)}\right.}
\end{equation}
At the same time, the presence inside a sample of a few scattering centers
(which are characterized by a mean-free path, $l$, $l>a$) inspires the
transition of the system from nonergodic (class $I$) to ergodic (class $II$)
state with the (inverse) relaxation time ($v_F$ is the Fermi velocity)
\begin{equation}
\frac 1\tau =(v_Fa){\left( \frac{\pi Ha}{4\phi _0}\right) }^2
\end{equation}
Turning to the HTS single crystals, let us consider dynamic (relaxation) and
equilibrium properties of the magnetization versus oxygen defect
concentration within the SG model. By analogy with the case of slightly
defected thin films, considered by de Gennes and Tinkham [18], we assume that
up to some critical value of oxygen deficiency, $\delta _g$, HTS single
crystals exhibit nonergodic (phase-coherent) behavior, while for
oxygen-defect concentration greater than $\delta _g$, the above-mentioned
coherence (within $CuO$ plane) is destroyed and crystal undergoes the phase
transition to ergodic (paracoherent) state where oxygen-rich superconducting
''grains'' are separated by oxygen-poor insulating boundaries so that there
is no superconducting path through a sample. It is worthwhile to mention that
the related problem of annealed Ising magnet on percolation clusters has been
recently considered by Kaufman and Touma [20]. Using the renorm-group method,
three phases on the corresponding phase diagram have been identified [20]:
percolating ferromagnetic, percolating paramagnetic, and nonpercolating
paramagnetic.

In view of Eqs.(10)-(13), the
equilibrium magnetization $M_{eq}(\delta ,T,H,H_1)$ is the limit
\begin{eqnarray}
M_{eq}(\delta ,T,H,H_1)&\equiv &\lim _{t\to \infty }M(t)\\ \nonumber
&=&M(\delta ,T,H,H_1)L^2(\delta ,T,H).
\end{eqnarray}
Here $L(\delta ,T,H)$ is the order parameter of the SG model which is
defined via the correlator $D(t)$ according to Eq.(13). To find the
long-time (low-frequency) behavior of the correlator $D(t)$ (and thus of the
magnetization $M(t)$), we need the equation of motion for the Josephson
pseudospins $S_i^{\pm }(t)$. An approximate (valid for $N\gg 1$) equation of
motion reads (see Appendix B) [7,12-16]
\begin{equation}
\dot S_i^{+}=\beta \Omega \sum_j^NJ_{ij}S_j^{+}
\end{equation}
Here $\Omega =2e^2R/\beta \hbar ^2N$ is a characteristic frequency of the JJ
network with $R$ being the resistance between grains in their normal state.
In the so-called ''mode-coupling approximation'' [21], $D(t)$ obeys the
self-consistent master equation (see Appendix B)
\begin{equation}
\frac{d^2D(t)}{dt^2}+\Omega ^2D(t)+\int\limits_0^tdt^{\prime }K(t-t^{\prime
})\frac{dD(t^{\prime })}{dt^{\prime }}=0,
\end{equation}
with $K(t)\equiv (1/N)\sum_{ij}K_{ij}(t)$ being a memory (feedback) kernel.
When there is no temporal correlations between grains (''paracoherent
state'') the memory kernel has a ''white noise'' form $K(t)\equiv
K_r(t)=2\Omega \delta (t)$, where $\delta (t)$ is the Dirac delta function.
In this case the master equation results in a Debye-like decay of
uncorrelated paracoherent state, namely $D(t)=exp(-t/\tau )$, where $1/\tau
=\Omega $. Such a situation is realized above some critical (phase-locking)
temperature $T_g$ when the coherent state within the JJ network is destroyed
completely, so that the order parameter $L\equiv 0$. Below $T_g$, the
situation changes drastically due to the superconducting correlations
occurring between grains. For $N\gg 1$, the coherent part of the memory
kernel, $K_{ij}^c(t)$, can be approximated by the ''current-current''
correlator $K_{ij}^c(t)\cong \overline{<\dot S_i^{+}(t)\dot S_j^{-}(0)>}$.
Taking into account the equation of motion (18), the memory kernel below $T_g
$ can be presented in the form (see Appendix B)
\begin{eqnarray}
K(t)&\equiv &K_r(t)+\frac 1N\sum_{ij}^NK_{ij}^c(t)\\ \nonumber
&=& 2\Omega \delta (t)+\Omega _{coh}^2(\delta ,T,H)D(t).
\end{eqnarray}
Here $\Omega _{coh}(\delta ,T,H)=\beta \Omega J(\delta ,T,H)$ and the field
dependence of the Josephson energy is defined as follows (see Appendix A)
\begin{eqnarray}
J(\delta ,T,H)&\equiv &\overline{J_{ij}(\delta ,T,H)}\\ \nonumber
&=&J(\delta ,T)\left( 1+\frac{H^2}{H_0^2}\right)^{-1}
\end{eqnarray}
In view of Eq.(13), a zero frequency ($t\rightarrow \infty $) solution of
the master Eq.(19) with the memory kernel (20) results in the nontrivial
order parameter for the intragranular JJ network [7,12-16]
\begin{equation}
L(\delta ,T,H)=1-\left( \frac{k_BT}{J(\delta ,T,H)}\right) ^2.
\end{equation}
The phase-locking temperature $T_g(\delta ,H)$, below which the ensemble of
grains undergoes the phase transition into the coherent state, is defined by
the equation $L(\delta ,T_g,H)=0$ which, due to Eq.(22), gives rise to
implicit equation, viz. $T_g(\delta ,H)=J(\delta ,T_g,H)/k_B$. The Josephson
energy depends on the temperature through the Ambegaokar-Baratoff relation,
which near the single grain superconducting temperature, $T_c$, reads $%
J(T)\approx J(0)(1-T/T_c)$. Assuming that for high magnetic fields (when
frustration is strong enough) $J(\delta ,0,H)\ll k_BT_c\leq J(\delta ,0,0)$,
we get finally $T_g(\delta ,H)\approx J(\delta ,0,H)/k_B$. As a result, the
order parameter $L=1-[T/T_g(\delta ,H)]^2$ gradually changes from $0$ at $%
T\geq T_g(\delta ,H)$ to $1$ at $T=0$, thus describing the continuous phase
transition.

By analogy with the critical (phase-locking) temperature $T_g(\delta
,H)$, we can introduce the critical field $H_g(\delta ,T)$ as the solution
of the equation $L(\delta ,T,H_g)=0$. In view of the field dependence of the
order parameter (see Eqs.(21) and (22)), the critical field reads
\begin{equation}
H_g(\delta ,T)=H_0\sqrt{1-T/T_g(\delta ,0)}.
\end{equation}
Taking into account the $\delta $ dependence of the phase-locking
temperature, $T_g(\delta ,H)\simeq T_g(0,H)(1-\delta )$, Eq.(23) results in
the following oxygen-deficiency behavior of the critical field
\begin{equation}
H_g(\delta ,T)=H_0\sqrt{\delta _g(T,0)-\delta }.
\end{equation}
Here we have introduced the critical oxygen deficiency, $\delta _g(T,H)$,
which is defined as the solution of the equation $L(\delta _g,T,H)=0$ and
has the form $\delta _g(T,H)=1-T/T_g(0,H)$. The physical meaning of this
critical parameter is as follows. For $\delta \geq \delta _g(T,H)$
oxygen-rich superconducting grains are separated by oxygen-poor insulating
boundaries so that there is no percolative path through the sample. Notice
that within the SG model, $H_g(\delta ,T)$ in fact plays the role of the
''phase boundary'' field $H_m(\delta ,T)$ discussed by Osofsky et al. [3].

It is important to mention that the correlator $D(t)$ follows a
simple Debye-like decay law only above $T_g(\delta ,H)$, i.e., when the
system of ''grains'' is in the ergodic state (see above). Below $T_g$ (where
the order parameter $L\neq 0$), relaxation of $D(t)$ (with $Im\{D(t)\}=0$)
can be presented in the form [13-16]
\begin{equation}
D(t)=L+(1-L)\widetilde{\Phi }(t).
\end{equation}
The relaxation function, $\widetilde{\Phi }(t)$, is supposed to be
normalized, viz.
\begin{equation}
\frac 1\tau \int\limits_0^\infty dt\widetilde{\Phi }(t)=1
\end{equation}
and obeys the following boundary conditions, $\widetilde{\Phi }(0)=1$ and $
\widetilde{\Phi }(\infty )=0$, i.e., $D(0)=1$. Of course, in principle, one
can find $D(t)$ as a numerical solution of the master equation. But it seems
more interesting to try and get some analytical results concerning the time
behavior of $D(t)$. It is natural then to consider a simple generalization
of the Debye law in the form of the so-called ''Kohlrausch stretched
exponential'' law
\begin{equation}
\widetilde{\Phi }(t)=\exp\left[ -\left( \frac t\tau \right)^\alpha \right],
\end{equation}
where $\alpha (\delta ,T,H)\leq 1$. Substitution of Eqs.(25)-(27) into
Eq.(19) with the kernel (20) results in an implicit equation on the power
exponent $\alpha (\delta ,T,H)$
\begin{equation}
\Gamma \left( 1+\frac 1\alpha \right) =1+\frac L{2(1-L)}.
\end{equation}
Here $\Gamma $ is the Gamma-function. Near $T_g(\delta ,H)$ the approximate
solution of Eq.(28) gives
\begin{equation}
\alpha (\delta ,T,H)\approx 1-L(\delta ,T,H).
\end{equation}
It is worthwhile to mention that for short-time limit, when $(t/\tau
)^\alpha \ll 1$, the above Kohlrausch law (27) leads to the non-logarithmic
relaxation law for magnetization (see Appendix B)
\begin{equation}
M(t)=M_{eq}[1-2s_\alpha log(t/\tau )+s_\alpha ^2log^2(t/\tau )].
\end{equation}
In this approximation, the $\alpha $-relaxation rate $s_\alpha (\delta ,T,H)$
is expressed via the Kohlrausch exponent $\alpha (\delta ,T,H)$ and the
order parameter $L(\delta ,T,H)$
\begin{equation}
s_\alpha (\delta ,T,H)=\left( \frac{1-L}L\right) \alpha .
\end{equation}
In turn, $s_\alpha (\delta ,T,H)$ is related to the activation energy $%
U(\delta ,T,H)$ as follows: $s_\alpha =k_BT/U$. Thus, in view of
Eqs.(29)-(31) the $\delta $, temperature, and field dependencies of the
activation energy in the JJ network are effective via the corresponding
dependencies of the order parameter $L(\delta ,T,H)$. Namely, near $%
T_g(\delta ,H)$ the activation energy reads $U(\delta ,T,H)\approx
k_BTL(\delta ,T,H)$. That is $U(\delta ,T,H)/2k_BT\approx 1-T/T_g(\delta
,H)\approx \delta _g(T,H)-\delta \approx 1-H/H_g(\delta ,T)$, in at least
qualitative agreement with what have been really observed in oxygen-depleted
HTS single crystals [4,5]. It is interesting to notice that expression
similar to our Eq.(30) has been used by Sengupta et al. [22] to describe a
non-logarithmic relaxation in HTS single crystals.

In view of the explicit dependence of the order parameter on the
oxygen deficiency, namely $L(\delta ,T,H)\approx 2(\delta _g(T,H)-\delta )$,
Eq.(29) describes the restoration of ergodic state in the system under the
study when oxygen deficiency $\delta $ reaches its critical value $\delta
_g(T,H)$. Indeed, when $\delta \rightarrow \delta _g(T,H)$, the order
parameter $L\rightarrow 0$, Kohlrausch exponent $\alpha \rightarrow 1$ (see
Eq.(29)) which means that relaxation becomes faster (formally, according to
Eq.(31) the logarithmic relaxation rate $s_\alpha \rightarrow \infty $) and
follows the ordinary Debye law (see Eq.(27)). At the same time, the
activation energy between ''grains'' declines (i.e., $U\rightarrow 0$) as $%
\delta \rightarrow \delta _g(T,H)$. To make our discussion more
quantitative, let us consider some estimates of the model parameters. Using
the experimental results for the ''phase boundary'' field $H_g(\delta ,T)$
obtained by Osofsky et al. [3] for fixed values of $\delta $ and $T$, namely $%
H_g(\delta =0.06,T=60K)\approx $ $1.5T$, $H_g(\delta =0.13,T=60K)\approx 1T$%
, and $H_g(\delta =0.13,T=70K)\approx 0.4T$, Eqs.(23) and (24) allow us to
get estimates for the phase-locking temperature $T_g(\delta ,H)$ and the
critical value of the oxygen deficiency $\delta _g(T,H)$. The result is: $%
T_g(\delta =0,H=0)\approx 75K$, $T_g(\delta =0.13,H=0)\approx 72K$, and $%
\delta _g(T=60K,H=0)\approx 0.21$. Finally, using the above results, Eq.(24)
brings about the following estimate for the characteristic Josephson field $%
H_0={\phi _0}/s\approx 5T$ which gives a reasonable value of oxygen ion
scattering cross section [3] $s\approx 4\times 10^{-16}m^2$. On the other
hand, making use of the above-obtained estimates we can estimate the value
of the activation energy $U(\delta ,T,H)\approx 2k_BT(1-$ $H/H_g(\delta ,T))$%
. For $H=1T$, $\delta =0.06$, and $T=60K$, we get $U/k_B\approx 40K$, which
reasonably agrees with the value deduced by Ossandon et al. [5] from $%
YBa_2Cu_3O_{7-\delta }$ single crystals measurements.

In summary, the oxygen defects induced phase transition from the
nonergodic (in nearly full-oxygenated crystals) to ergodic (in
highly-deoxygenated crystals) state in HTS oxygen-depleted crystals has
been considered within the superconductive glass model. Both dynamic
(relaxation) and equilibrium properties of the model magnetization were
found to correlate quite reasonably with some experimental data on
deoxygenated HTS.

\appendix
\section{}

To get Eqs.(10) and (11) for the relaxation of magnetization, we
have to calculate the relaxation function $\Phi (t)$. Using the so-called
random-field approximation for quenched disordered systems [10,16,23], which
allows to decouple the averaging of the ''grain distribution'' (represented
by the ''scattering potentials'' $J_{ij}$) from the ''carriers'' (or
Josephson pseudospins), i.e., assuming that $\overline{A(r_i)B(r_j)}\cong
\overline{A(r_i)}\cdot \overline{B(r_j)}$, we obtain from Eqs.(2)-(4), and
(9)
\begin{eqnarray}
\Phi (t) &\cong &-\frac{e^2\beta J^2}{\hbar ^2}\sum_{ijkl}^N
\{\overline{<S_i^{+}(t)S_j^{-}(t)S_k^{+}(0)S_l^{-}(0)>}\\ \nonumber
& \times & \overline{\exp [i(A_{ij}+A_{kl})]{(x_iy_j-x_jy_i)(x_ky_l-x_ly_k)}}\\ \nonumber
& -& \overline{\exp[i(A_{ij}-A_{kl})]{(x_iy_j-x_jy_i)(x_ky_l-x_ly_k)}}\\ \nonumber
& \times &\overline{<S_i^{+}(t)S_j^{-}(t)S_k^{-}(0)S_l^{+}(0)>}\}+h.c.
\end{eqnarray}
To proceed further, we have to calculate the 4-spin correlators appeared in
the rhs of the above equation. Taking into account only pair correlations,
namely assuming that (within mean-field approximation [10,16,23], when $N\gg 1$%
), e.g., $\overline{<S_i^{+}(t)S_j^{-}(t)S_k^{+}(0)S_l^{-}(0)>}\cong
\overline{<S_i^{+}(t)S_l^{-}(0)>}\cdot \overline{<S_j^{-}(t)S_k^{+}(0)>}$,
we can rewrite Eq.(A1) as follows
\begin{eqnarray}
\Phi (t) & \cong &
\frac{\phi _0^2e^2\beta J^2}{\hbar ^2}\sum_{ijkl}^N
\left[ \frac \partial {\partial H}\overline{e^{iA_{ij}}}\right] \left[
\frac \partial {\partial H}\overline{e^{iA_{kl}}}\right]\\ \nonumber
& \times & \{D_{il}(t)D_{jk}^{*}(t)+D_{ik}(t)D_{jl}^{*}(t)\}+h.c.
\end{eqnarray}
Here we have introduced the spin-spin correlator (cf. Eq.(12))
\begin{equation}
D_{ij}(t)=\overline{<S_i^{+}(t)S_j^{-}(0)>},
\end{equation}
and made use of the fact that due to Eq.(3)
\begin{equation}
\overline{{(x_iy_j-x_jy_i)\exp (iA_{ij})}}=\left( \frac{i\phi _0}\pi \right)
\frac \partial {\partial H}\overline{\exp (iA_{ij})}.
\end{equation}
The frustration field dependence of magnetization essentially depends on the
choice of a random distribution function $P(\vec r_i)$ as well as on the
type of disorder [9-15]. To obtain the explicit form of the field dependence
of magnetization given by Eq.(11), we have assumed, for simplicity, a
site-type positional disorder allowing for weak displacements of the grain
sites from their positions of the original $2D$ lattice, i.e., within a
radius $d$ the new position is chosen randomly according to the normalized
separable Gaussian distribution function $P(\vec r_i)=P(x_i)P(y_i)$, where
\begin{equation}
P(x)=\frac 1{\sqrt{2\pi d^2}}exp\left( -\frac{x^2}{2d^2}\right).
\end{equation}
Using the above distribution function, we can calculate the configurational
averages appeared in Eq.(A2). In particular, the average value of the
Josephson energy (see Eq.(2)) reads
\begin{equation}
J(\delta ,T,H)\equiv \overline{J_{ij}(\delta ,T,H)}=J(\delta ,T)\overline{%
\exp (iA_{ij})},
\end{equation}
where
\begin{eqnarray}
\overline{e^{iA_{ij}}}&\equiv &\int\limits_{-\infty }^{+\infty }d\vec
r_id\vec r_jP(\vec r_i)P(\vec r_j)e^{(i\pi H/\phi _0){(x_iy_j-x_jy_i)}}\\ \nonumber
&=&\left( 1+\frac{H^2}{H_0^2}\right) ^{-1}.
\end{eqnarray}
Here $d\vec r_i=dx_idy_i$, and $H_0={\phi _0}/\pi d^2$. Finally, taking into
account Eqs.(A3)-(A7), we arrive to Eqs.(10) and (11) for magnetization $%
M(t) $.

\section{}

By accounting for the Kirchhoff law, $\sum_iI_{ij}=0$, for the total
Josephson currents, $I_{ij}=I_{ij}^s+I_{ij}^n$, where the superconducting
current, $I_{ij}^s$, is given by Eq.(4) and $I_{ij}^n=(\hbar /2eR)(d\phi
_{ij}/dt)$ is a normal current with $R$ being the resistance between grains
in their normal state, the approximate (valid for $N\gg 1$) equation of
motion for the superconducting phase reads [10,14]
\begin{equation}
\frac{\hbar N}{2eR}\frac{d\phi _i}{dt}+\frac{2eJ}\hbar \sum_j^N\sin \phi
_{ij}=0
\end{equation}
Taking into account that $(d/dt)e^{+i\phi _i}=i(d\phi _i/dt)e^{+i\phi _i}$,
the pseudospin representation (with $S_i^{+}=e^{+i\phi _i}$) brings
about the approximate equation of motion (18) for Josephson pseudospins.

In the so-called ''mode-coupling approximation'' [21], which is based
on the Mori-like projection technique [23,24], the self-consistent master
equation on the isothermal correlation function $D(t)$ can be constructed.
Let us introduce the Laplace transform
\begin{equation}
D_{ij}(z)\equiv i\int\limits_0^{+\infty }dte^{izt}D_{ij}(t)
\end{equation}
Then the continued fraction expansion for $D_{ij}(z)$ leads to the
expression [14,21]
\begin{equation}
D_q(z)=-\left( z-\frac{\Omega ^2}{z+K_q(z)}\right) ^{-1},
\end{equation}
where
\begin{equation}
D_q(z)=\frac 1N\sum_{jk}^Ne^{iq(j-k)}D_{jk}(z).
\end{equation}
Here $\Omega =2e^2k_BTR/\hbar ^2N$ is a characteristic frequency of the JJ
network. Alternatively, using inverse Laplace transform, Eq.(B3) can be cast
into the self-consistent master equation (Eq.(19)). Using the mode-coupling
approximation scheme [14,21], the coherent part of the memory kernel can be
represented by a set of ''current-current'' correlators
\begin{equation}
K_{ij}^c(t)\simeq \overline{<\dot S_i^{+}(t)\dot S_j^{-}(0)>}+\Omega ^2\overline{%
<\ddot S_i^{+}(t)\ddot S_j^{-}(0)>}
\end{equation}
Since $\dot S_i^{+}\propto \Omega $ (see Eq.(18)), due to a rather strong
dependence of the characteristic frequency $\Omega $ on the number of grains
($\Omega \propto 1/N$), we can restrict ourselves to a linear approximation,
$K_{ij}^c(t)\cong \overline{<\dot S_i^{+}(t)\dot S_j^{-}(0)>}$, assuming
that $N\gg 1$. Taking into account the equation of motion (18), $%
K_c(t)\equiv (1/N)\sum_{ij}K_{ij}^c(t)$ can be presented in the form
\begin{eqnarray}
K_c(t)&\cong &\frac{\beta ^2\Omega ^2}N\sum_{ij}^N\sum_{kl}^N\overline{%
J_{ik}J_{jl}}\cdot \overline{<S_k^{+}(t)S_l^{-}(0)>}\\ \nonumber
&=&\beta ^2\Omega ^2J^2(\delta ,T,H)D(t).
\end{eqnarray}
To obtain the above equation, we have used Eqs.(A3), (A6), and (A7) together
with the decoupling approximations discussed in Appendix A. Using Eq.(B6),
we finally arrive at Eq.(20) for the total memory kernel below $T_g$.

To get the non-logarithmic relaxation law (30) for magnetization,
let us rewrite Eq.(10) taking into account Eqs.(25) and (27) as follows
\begin{equation}
\log \left( 1-Z\right) =\log \left[ 1-exp\left( -\left( \frac t\tau \right)
^\alpha \right) \right]
\end{equation}
Here $Z\equiv (\sqrt{M(t)}-\sqrt{M_{eq}})/(\sqrt{M_0}-\sqrt{M_{eq}})$%
, $M_0\equiv M(\delta ,T,H,H_1);$ $M_{eq}$ and $M(\delta ,T,H,H_1)$ are
given by Eqs.(17) and (10), respectively. For short-time limit, when $%
(t/\tau )^\alpha \ll 1$, and $Z\ll 1$, we can expand both sides of Eq.(B7)
and get in the linear approximation $\sqrt{M(t)}-\sqrt{M_{eq}}\cong -(\sqrt{%
M_0}-\sqrt{M_{eq}})\alpha \log (t/\tau )$. Finally, taking into account that
$M_{eq}=M_0L^2$ (see Eq.(17)), we find (cf.Eq.(30)) $M(t)=M_{eq}[1-s_\alpha
log(t/\tau )]^2$, where $s_\alpha \equiv \alpha (\sqrt{M_0}-\sqrt{M_{eq}})/
\sqrt{M_{eq}}=\alpha (1-L)/L$.


\begin{thebibliography}{99}
\bibitem{ref1}  M. Daeumling, J.M. Seuntjens, and D.C. Larbalestier,
{\it Nature (London)} {\bf 346}, 332 (1990).

\bibitem{ref2}  S. Sergeenkov, {\it J. Superconduct.} {\bf 4}, 431 (1991).

\bibitem{ref3}  M.S. Osofsky, J.L. Cohn, E.F. Skelton, M.M. Miller,
R.J. Soulen, Jr., S.A. Wolf, and T.A. Vanderah, {\it Phys.Rev.B} {\bf 45}, 4916
(1992).

\bibitem{ref4}  J.G. Ossandon, J.R. Thompson, D.K. Christen, B.C. Sales,
H.R. Kerchner, J.O. Thomson, Y.R. Sun, K.W. Lay, and J.E. Tkaczyk, {\it %
Phys.Rev.B} {\bf 45}, 12534 (1992).

\bibitem{ref5}  J.G. Ossandon, J.R. Thompson, D.K. Christen, B.C. Sales,
Y.R. Sun, and K.W. Lay, {\it Phys.Rev.B} {\bf 46}, 3050 (1992).

\bibitem{ref6}  M. Wolf, J. Gleitzmann and W. Gey, {\it Phys.Rev.B} {\bf %
47}, 8381 (1993).

\bibitem{ref7}  S. Sergeenkov and M. Ausloos, {\it Phys.Rev.B} {\bf 47}, 14476
(1993).

\bibitem{ref8}  M. Kaufman, Physica A {\bf 177}, 523 (1991).

\bibitem{ref9}  C. Ebner and D. Stroud, {\it Phys.Rev.B} {\bf 31}, 165
(1985).

\bibitem{ref10}  V.M. Vinokur, L.B. Ioffe, A.I. Larkin, and M.V. Feigelman,
{\it Sov. Phys. JETP} {\bf 66}, 198 (1987).

\bibitem{ref11}  I. Morgenstern, K.A. M\"uller, and J.G. Bednorz, {\it %
Z.Phys.B} {\bf 69}, 33 (1987).

\bibitem{ref12}  J. Choi and J.V. Jose, {\it Phys.Rev.Lett.} {\bf 62},
320 (1989).

\bibitem{ref13}  S. Sergeenkov and M. Ausloos, {\it Phys.Rev. B }{\bf 48},
604 (1993).

\bibitem{ref14}  S. Sergeenkov, Physica C {\bf 205}, 1 (1993).

\bibitem{ref15}  S. Sergeenkov and M. Ausloos, {\it Phys.Rev.B} {\bf 46%
}, 14223 (1992).

\bibitem{ref16}  S. Sergeenkov and M. Ausloos, {\it Phys.Rev.B }{\bf %
48}, 4188 (1993).

\bibitem{ref17}  R. Kubo, {\it J.Phys.Soc.Jpn.} {\bf 12}, 570 (1957).

\bibitem{ref18}  P.G. de Gennes, {\it Superconductivity of Metals
and Alloys} (W.A. Benjamin, Inc., New York, 1966), Chapter 8.

\bibitem{ref19}  P.G. de Gennes and M. Tinkham, {\it Physics} {\bf 1},
107 (1964).

\bibitem{ref20}  M. Kaufman and J.E. Touma, {\it Phys.Rev.B }{\bf %
49}, 9583 (1994).

\bibitem{ref21}  W. Gotze and L. Sjogren, {\it J.Phys.C} {\bf 17},
5759 (1984).

\bibitem{ref22}  S. Sengupta, D. Shi, Z. Wang, M.E. Smith, and
P.J. McGinn, {\it Phys.Rev. B }{\bf 47}, 5165 (1993).

\bibitem{ref23}  P.L. Taylor, {\it A Quantum Approach to the Solid
State }(Prentice-Hall, Englewood Cliffs, NJ, 1970), Chapters 2 and 3.

\bibitem{ref24}  D. Forster, {\it Hydrodynamic Fluctuations, Broken
Symmetry and Correlation Functions }(W.A.Benjamin, Reading, New York, 1975).
\end{thebibliography}
\end{document}